\documentclass[aps,jcp,showpacs,showkeys,preprint, endfloats]{revtex4}

\usepackage{amsmath}
\usepackage{amsfonts}            
\usepackage{amssymb}

\usepackage{mathrsfs}

\usepackage{eufrak}

\usepackage{graphicx}
\usepackage{epstopdf}

\usepackage{dcolumn}
\usepackage{bm}

\makeatletter
\def\@dotsep{4.5}
\makeatother

\begin{document}

\title{An analytic model of rotationally inelastic collisions of polar molecules in electric fields}

\author{Mikhail Lemeshko}

\author{Bretislav Friedrich}

\affiliation{%
Fritz-Haber-Institut der Max-Planck-Gesellschaft, Faradayweg 4-6, D-14195 Berlin, Germany
}%

\date{\today}

\begin{abstract}

We present an analytic model of thermal state-to-state rotationally inelastic collisions of polar molecules in electric fields. The model is based on the Fraunhofer scattering of matter waves and requires Legendre moments characterizing the ``shape" of the target in the body-fixed frame as its input. The electric field orients the target in the space-fixed frame and thereby effects a striking alteration of the dynamical observables: both the phase and amplitude of the oscillations in the partial differential cross sections undergo characteristic field-dependent changes that transgress into the partial integral cross sections. As the cross sections can be evaluated for a field applied parallel or perpendicular to the relative velocity, the model also offers predictions about steric asymmetry. We exemplify the field-dependent quantum collision dynamics with the behavior of the Ne-OCS($^{1}\Sigma$) and Ar-NO($^2\Pi$) systems. A comparison with the close-coupling calculations available for the latter system [Chem. Phys. Lett. \textbf{313}, 491 (1999)] demonstrates the model's ability to qualitatively explain the field dependence of all the scattering features observed.

\end{abstract}

\maketitle

\section{Introduction}
Collisions of molecules in electric, magnetic, or radiative fields are nearly ubiquitous in nature as well as in the laboratory. Molecules colliding in the Earth's atmosphere or in interstellar space are commonly subjected to magnetic and radiative fields; in the laboratory, collisions in fields appear with particular prominence in stereodynamics~\cite{epjdstereospecial}, coherent control~\cite{cohcontrol}, and molecular trapping and cooling~\cite{epjdcoolspecial}. Molecular collisions in fields have been the subject of a number of theoretical studies, recently reviewed, e.g., in Ref.~\cite{Krems04}. However, analytic models of such collisions are scarce, and limited to the collision regime near the Wigner limit, see, e.g., ref.~\cite{sadeghpour2000}. Here we present an analytic model of state-to-state rotationally inelastic collisions of atoms with polar molecules in electric fields. The model is based on the Fraunhofer scattering of matter waves and is applicable to collisions at thermal and hyperthermal energies. We develop the model for the collisions of closed-shell atoms with $^1\Sigma$ (linear) or $^2\Pi$ (symmetric-top equivalent) molecules, and compare it with the close-coupling calculations of van Leuken \emph{et al.}~\cite{vanLeuken96}, available for the latter system.

The field-free Fraunhofer model was developed by Drozdov~\cite{Drozdov} and generalized by Blair~\cite{Blair} in the late 1950s  to treat inelastic nuclear scattering. The model provided a much-sought explanation of the experimentally observed phase shifts between oscillations in the elastic and inelastic differential cross sections for the scattering of protons or $\alpha$ particles by medium-sized nuclei, later referred to as the ``Blair phase rule." In 1984, the field-free Fraunhofer model was adapted by Faubel~\cite{Faubel} to account for rotationally inelastic thermal collisions between helium atoms and N$_2$ and CH$_4$ molecules. 

In this paper, we extend the model to include the effects of an electrostatic field on rotationally inelastic scattering of polar molecules by atoms. Within the model, these effects arise due to the orientation of the polar molecules in the space-fixed frame and the concomitant relaxation of the parity selection rule. Although the model -- in both its field-free and field-dependent incarnation -- is only semiquantitative,  it readily explains all the features found in the state-to-state differential and integral cross sections and in their dependence on the strength and direction of the electrostatic field. These features include the phases of the angular oscillations in the differential cross section and their characteristic variation as a function of the electric field. 

In Section~\ref{sec:FraunApprox}, we prepare the soil by introducing the field-free Fraunhofer model of matter-wave scattering. In Sections~\ref{sec:FraunSigma} and~\ref{sec:Fraun2Pi}, we extend the Fraunhofer model to account for scattering of polar molecules in electric fields. In Section~\ref{sec:FraunSigma}, we work out closed-form expressions for the partial and total differential and integral cross sections and the steric asymmetry for collisions between closed-shell atoms and polar $^1\Sigma$ molecules, and apply them to the $\text{Ne-OCS}(^1\Sigma, J=0 \to J')$ collision system. Section~\ref{sec:Fraun2Pi} develops the theory for collisions between closed-shell atoms and polar~$^2\Pi$ molecules in electric fields and exemplifies the results by treating the $\text{Ar-NO }(^2\Pi, J=\frac{1}{2} \to J')$ collision system. The main conclusions of this work are summarized in Section~\ref{sec:conclusions}.

\section{The Fraunhofer model of field-free scattering}
\label{sec:FraunApprox}

We first describe the Fraunhofer model of field-free scattering and discuss its validity. The model is based on two approximations:

(i) The energy \textit{sudden approximation}, which represents the amplitude 
\begin{equation}
	\label{InelAmplSudden}
	f_{\mathfrak{i} \to \mathfrak{f}} (\vartheta) = \langle   \mathfrak{f} \vert f_{\text{el}}(\vartheta) \vert \mathfrak{i} \rangle
\end{equation}
for scattering into an angle $\vartheta$ from an initial, $\vert \mathfrak{i} \rangle$, to a final,  $\vert \mathfrak{f} \rangle$, state in terms of the elastic scattering amplitude, $f_{\text{el}}(\vartheta)$, at fixed values of the internal coordinates. 
The energy sudden approximation is well justified when the collision energy exceeds the spacing of the internal states, $E_{\text{coll}}\gg\Delta E_{\text{int}}$ \cite{Blair}, \cite{Faubel}. 

\begin{figure*}[htbp]
\centering\includegraphics[width=6cm]{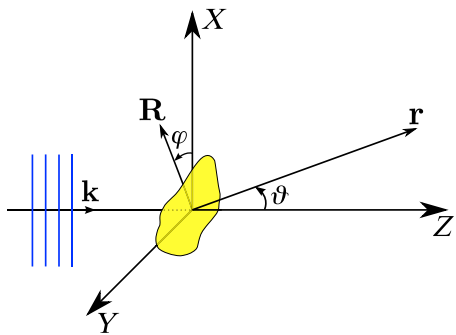}
\caption{Schematic of Fraunhofer diffraction by an impenetrable, sharp-edged obstacle as observed at a point of radius vector $\textbf{r}(X,Z)$ from the obstacle. Relevant is the shape of the obstacle in the $XY$ plane, perpendicular to the initial wave vector, $\mathbf{k}$, itself directed along the $Z$-axis of the space-fixed system $XYZ$. The angle $\varphi$ is the polar angle of the radius vector $\textbf{R}$ which traces the shape of the obstacle in the $X,Y$ plane and $\vartheta$ is the scattering angle. See text.}\label{fig:fraunhofer}
\end{figure*}

(ii) The elastic scattering amplitude $f_{\text{el}}(\vartheta)$ in Eq.~(\ref{InelAmplSudden}) is replaced by the amplitude for \textit{Fraunhofer diffraction} by an impenetrable, sharp-edged obstacle as observed at a point of radiusvector $\textbf{r}$ from the obstacle, see Fig.~\ref{fig:fraunhofer}. 
In its simplest form, the Fraunhofer diffraction amplitude is given by the integral
\begin{equation}
	\label{FraunAmpl}
	f {(\vartheta)} \approx \int e^{-i k R \vartheta \cos \varphi} d R
\end{equation}
where $\varphi$ is the polar angle of the radius vector $\textbf{R}$ which traces the shape of the obstacle, $R\equiv|\mathbf R|$, and $k\equiv|\mathbf k|$ with $\mathbf k$ the initial wave vector. Relevant is the shape of the obstacle in the space-fixed $XY$ plane, perpendicular to $\mathbf{k}$, itself directed along the space-fixed $Z$-axis, cf. Fig.~\ref{fig:fraunhofer}. The major approximation made in deriving Eq.~(\ref{FraunAmpl}) consists in neglecting terms non-linear in $\textbf{r}$. We note that the notion of a sharp-edged obstacle comes close to the rigid shell approximation. The latter has been widely used in classical~\cite{Beck79},~\cite{IchimuraNakamura},~\cite{Marks_ellips}, quantum~\cite{Bosanac}, and quasi-quantum~\cite{Stolte} treatments of field-free molecular collisions, where the collision energy by far exceeds the depth of any potential energy well.

In optics, Fraunhofer (i.e., far-field) diffraction~\cite{BornWolf} occurs when the Fresnel number is small,
\begin{equation}
	\label{FresnelNumber}
	\mathscr{F} \equiv \frac{a^2}{r \lambda} \ll 1
\end{equation}
Here $a$ is the dimension of the obstacle, $r\equiv|\textbf{r}|$ is the distance from the obstacle to the observer, and $\lambda$ is the wavelength, cf. Fig.~\ref{fig:fraunhofer}. Condition~(\ref{FresnelNumber}) is well satisfied for nuclear scattering at MeV collision energies as well as for molecular collisions at thermal and hyperthermal energies. In the latter case, inequality~(\ref{FresnelNumber}) is fulfilled due to the compensation of the larger molecular size $a$ by a larger de~Broglie wavelength $\lambda$ pertaining to thermal molecular velocities. 

We note that the Fraunhofer scattering amplitude, Eq.~(\ref{FraunAmpl}), is quite similar to the amplitude for Born scattering~\cite{LLIII}. Either amplitude is a Fourier transform of the target's spatial characteristic -- either its shape or its potential. Both the Fraunhofer and Born amplitudes comprise averages of the phase factor, $\exp (i \textbf{kR})$, over the target's surface or volume~\cite{Blair59}.

For nearly-circular targets, with a boundary $R (\varphi) = R_0 +\delta(\varphi)$ in the $XY$ plane, the Fraunhofer integral of Eq.~(\ref{FraunAmpl}) can be evaluated and expanded in a power series in the deformation $\delta(\varphi)$,
\begin{equation}
	\label{AmplitudeExpansion}
	f_{}  {(\vartheta)}  = f_0 (\vartheta) + f_1 (\vartheta,\delta) + f_2(\vartheta,\delta^2)+\cdots
\end{equation}
with $f_0(\vartheta)$ the amplitude for scattering by a disk of radius $R_0$
\begin{equation}
	\label{AmplSphere}
	f_0 (\vartheta) = i (k R_0^2) \frac{J_1 (k R_0 \vartheta)}{(k R_0 \vartheta)}
\end{equation}
and $f_1$ the lowest-order anisotropic amplitude,
\begin{equation}
	\label{AmplFirstOrder}
	f_1(\vartheta) = \frac{i k}{2 \pi} \int_{0}^{2 \pi} \delta(\varphi) e^{- i (k R_0 \vartheta) \cos \varphi} d\varphi
\end{equation}
where $J_1$ is a Bessel function of the first kind. Both Eqs.~(\ref{AmplSphere}) and~(\ref {AmplFirstOrder}) are applicable at small values of $\vartheta \lesssim 30^{\circ}$, i.e., within the validity of the approximation $\sin \vartheta \approx \vartheta$. 

A key step required to maintain the analyticity of the Fraunhofer scattering amplitude, Eq.~(\ref{AmplFirstOrder}),  is to present the shape of the atom-linear molecule potential in terms of a series in spherical harmonics,

\begin{equation}
	\label{RhoExpBF}
	R^{\flat} (\theta^{\flat}, \phi^{\flat}) = \sum_{\kappa \nu} \Xi_{\kappa \nu} Y_{\kappa \nu} (\theta^{\flat}, \varphi^{\flat})
\end{equation}
with $\Xi_{\kappa \nu}$ the Legendre moments. The polar and azimuthal angles $\theta^{\flat}$ and $\varphi^{\flat}$ pertain to the body-fixed frame, defined, e.g., by the target's principal axes of inertia. However, what matters is the target's shape in the space fixed frame, see Fig.~\ref{fig:fraunhofer}, which is given by
\begin{equation}
	\label{RhoExpSpaceFixed}
	R (\alpha, \beta, \gamma ; \theta, \varphi) = \sum_{\kappa \nu \rho} \Xi_{\kappa \nu} \mathscr{D}_{\rho \nu}^{\kappa} (\alpha \beta \gamma) Y_{\kappa \rho} (\theta, \varphi)
\end{equation}
where $(\alpha,\beta,\gamma)$ are the Euler angles through which the body-fixed frame is rotated relative to the space-fixed frame, $(\theta, \varphi)$ are the polar and azimuthal angles in the space-fixed frame, and $\mathscr{D}_{\rho \nu}^{\kappa} (\alpha \beta \gamma)$ are the Wigner rotation matrices. Clearly, the term with  $\kappa=0$ corresponds to a disk of radius $R_0$,
\begin{equation}
	\label{R0viab}
	R_0 \approx \frac{\Xi_{00}}{\sqrt{4\pi}}
\end{equation}
Since of relevance is the shape of the target in the $XY$ plane, we set $\theta=\frac{\pi}{2}$ in Eq.~(\ref{RhoExpSpaceFixed}). As a result,
\begin{equation}
\label{deltaphi}
	\delta(\varphi)=R (\alpha, \beta, \gamma ; \tfrac{\pi}{2}, \varphi)-R_0=R (\varphi) - R_0=\underset{\kappa \neq 0 }{ \sum_{\kappa \nu \rho}} \Xi_{\kappa \nu} \mathscr{D}_{\rho \nu}^{\kappa} (\alpha \beta \gamma) Y_{\kappa \rho} (\tfrac{\pi}{2}, \varphi)
\end{equation}
By substituting from Eq.~(\ref{deltaphi}) into Eq.~(\ref{AmplFirstOrder}) and evaluating the integral, we obtain the following expression for the first-order scattering amplitude,
\begin{equation}
	\label{AmplFirstOrderFinal}
	f_1(\alpha,\beta,\gamma ; \vartheta) =  \frac{i k R_0}{2 \pi}  \underset{\kappa \neq 0 } {\sum_{\kappa \nu \rho}}  \Xi_{\kappa \nu} \mathscr{D}_{\rho \nu}^{\kappa} (\alpha \beta \gamma) F_{\kappa \rho} J_{\vert \rho \vert} (k R_0 \vartheta)
\end{equation}
with $F_{\kappa \rho}$ defined by
\begin{equation}
	\label{Flamnu}
	F_{\kappa \rho} = \left \{ 	\begin{array}{ccl}
		(-1)^{\rho} 2\pi \left( \frac{2\kappa+1}{4\pi} \right)^{\frac{1}{2}} (-i)^{\kappa} \frac{\sqrt{(\kappa+\rho)! (\kappa-\rho)! }}{(\kappa+\rho)!! (\kappa-\rho)!! }  &    &    \textrm{ for $\kappa+\rho$ ~even~and~ $\kappa \ge \rho$} 
		\\ \\						0     &    &     \textrm{ elsewhere} 
	\end{array}    \right .
\end{equation}
For negative values of $\rho$, the factor $(-i)^{\kappa}$ is to be replaced by $i^{\kappa}$.
Finally, by making use of Eq.~(\ref {InelAmplSudden}), we obtain the inelastic scattering amplitude as
\begin{equation}
	\label{InelAmplExpress}
	f_{\mathfrak{i} \to \mathfrak{f}} (\vartheta) \approx \langle \mathfrak{f} \vert f_0 + f_1 \vert \mathfrak{i} \rangle = \langle \mathfrak{f} \vert f_1 \vert \mathfrak{i} \rangle =  \frac{i k R_0}{2 \pi}  \underset{\kappa+\rho~\textrm{even}}{\underset{\kappa \neq 0 } {\sum_{\kappa \nu \rho}}} \Xi_{\kappa \nu} \langle \mathfrak{f} \vert \mathscr{D}_{\rho \nu}^{\kappa} \vert \mathfrak{i} \rangle F_{\kappa \rho} J_{\vert \rho \vert} (k R_0 \vartheta)
\end{equation}


\section{The Fraunhofer model of rotationally inelastic scattering of polar $^1\Sigma$ molecules by closed-shell atoms in an electrostatic field}
\label{sec:FraunSigma}

\subsection{The field-dependent scattering amplitude}
\label{sec:Fraun1Sigma}

When a polar $^1\Sigma$ molecule enters an electrostatic field, its rotational states undergo \textit {hybridization} (coherent linear superposition), induced by the interaction of the molecule's body-fixed electric dipole moment, $\boldsymbol{\mu }$, with the electric field, $\boldsymbol{\varepsilon}$~\cite{Loesch},~\cite{FriHer}. Because of the cylindrical symmetry about the electric field vector, the permanent-dipole interaction couples free-rotor basis states, $\vert J, M \rangle$, with a fixed value of the good quantum number, $M$,  but a range of $J$'s. Thus the hybrid wavefunctions take the general form
\begin{equation}
	\label{PendularStateGeneral}
	\vert \tilde{J}, M, \omega \rangle = \sum_{J} a_{J M}^{\tilde{J}} (\omega) \vert J, M \rangle
\end{equation}
where the expansion coefficients $a_{J M}^{\tilde{J}}$ depend solely on a dimensionless interaction parameter, 
\begin{equation}
	\label{omegapar}
	\omega \equiv \mu \varepsilon /B
\end{equation}
which measures the maximum potential energy, $\mu \varepsilon$, of the dipole in terms of the molecule's rotational constant, $B$. The symbol $\tilde{J}$ denotes the nominal value of $J$ that pertains to the field-free rotational state which adiabatically correlates with the hybrid state,
\begin{equation}
	\label{PendularToFreeRotor}
	\vert \tilde{J}, M, \omega \to 0 \rangle \to \vert  J, M \rangle
\end{equation}
and $\mu \equiv |\boldsymbol{\mu }|$, $\varepsilon \equiv |\boldsymbol{\varepsilon }|$.

In order to account for an arbitrary direction of the electric field with respect to the initial wave vector $\mathbf{k}$, we introduce a field-fixed coordinate system $X^{\sharp} Y^{\sharp}Z^{\sharp}$, whose $Z^{\sharp}$-axis is defined by the direction of the electric field vector $\boldsymbol{\varepsilon }$. The free-rotor states are thus given by spherical harmonics whose arguments are the angles $\theta^{\sharp}$ and $\varphi^{\sharp}$ in the field-fixed frame,
\begin{equation}
	\label{FreeRotorFieldFixed}
	\vert J, M \rangle = Y_{J M}(\theta^{\sharp},\varphi^{\sharp})
\end{equation}

\begin{figure*}[htbp]
\centering\includegraphics[width=15.5cm]{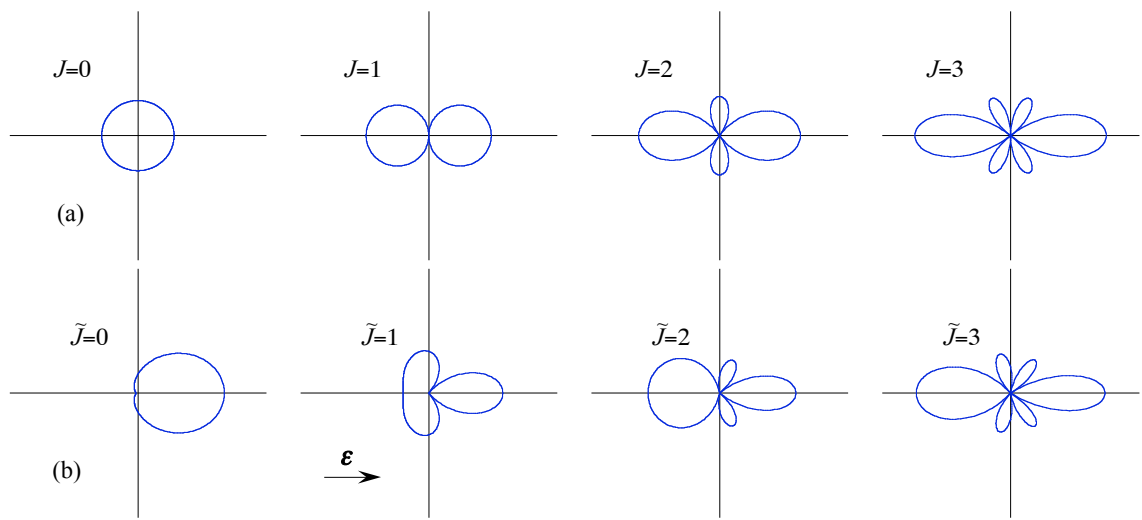}
\caption{A comparison of the moduli of the free rotor wavefunctions $\bigl | \vert J, M=0 \rangle \bigr |$, panel (a), with the moduli of the pendular wavefunctions $\bigl | \vert \tilde{J}, M=0, \omega=5 \rangle  \bigr |$, panel (b). Also shown is the direction of the electric field vector, $\boldsymbol{\varepsilon}$.}
\label{fig:pendstate}
\end{figure*}

Apart from possessing a \textit{sui generis} energy level pattern, the $\vert \tilde{J}, M, \omega \rangle$ eigenstates have an indefinite (mixed) parity and are directional, exhibiting a varying degree of orientation, which depends on the values of  $\tilde{J}$, $M$, and $\omega$. In the oriented states, the body-fixed dipole (and thus the internuclear axis) librates about the field direction like a pendulum, and so the hybrid states are referred to as \textit{pendular}. It is the directionality of the pendular states along with their mixed parity that enters the field-dependent Fraunhofer model and distinguishes it from the field-free model, which assumes an isotropic distribution of the molecular axes and a definite parity. The directional properties of pendular states are exemplified in Figure Fig.~\ref{fig:pendstate}, which shows polar diagrams of the field-free and pendular wave functions at $\omega=5$.

Hence the scattering process in the field comprises the following steps:  A molecule in a free-rotor state  $\vert J, M \rangle$ enters adiabatically the field where it is transformed into a pendular state $\vert \tilde{J}, M, \omega \rangle$. This pendular state may be changed by the collision in the field into another pendular state, $\vert \tilde{J'}, M', \omega \rangle$. As the molecule leaves the field, the latter pendular state is adiabatically transformed into a free-rotor state $\vert J', M' \rangle$. Thus the net result is, in general,  a rotationally inelastic collision,  $\vert J, M \rangle \rightarrow \vert J', M' \rangle$.

In order to be able to apply Eq.~(\ref{InelAmplExpress}) to collisions in the electrostatic field,  we have to transform Eq.~(\ref{PendularStateGeneral}) to the space-fixed frame $XYZ$. If the electric field vector is specified by the Euler angles $(\varphi_{\varepsilon},\theta_{\varepsilon},0)$ in the $XYZ$ frame, the initial and final pendular states take the form
\begin{align}
	\label{PendularStateASpFix}
	\vert \mathfrak{i} \rangle  \equiv \vert \tilde{J}, M, \omega \rangle & =  \sum_{J} a_{J M}^{\tilde{J}} (\omega)\sum_{\xi} \mathscr{D}_{\xi M}^{J} (\varphi_{\varepsilon},\theta_{\varepsilon},0)  Y_{J \xi} (\theta,\varphi) \\
	\label{PendularStateBSpFix}
	\langle \mathfrak{f} \vert  \equiv \langle \tilde{J'}, M', \omega \vert & =  \sum_{J'} b_{J' M'}^{\tilde{J'} \ast} (\omega) \sum_{\xi'} \mathscr{D}_{\xi' M'}^{J' \ast} (\varphi_{\varepsilon},\theta_{\varepsilon},0)  Y_{J' \xi'}^{\ast} (\theta,\varphi)	
\end{align}
which is seen to depend solely on the angles $\theta$ and $\varphi$ (and not the angles $\theta^{\sharp}$ and $\varphi^{\sharp}$ pertaining to the field-fixed frame).

On substituting from Eqs.~(\ref{PendularStateASpFix}) and~(\ref{PendularStateBSpFix}) into Eq.~(\ref{InelAmplExpress}) and its integration, we obtain a general expression for the Fraunhofer scattering amplitude in the field,
\begin{multline}
	\label{ScatAmplArbField}	
	f_{\mathfrak{i} \to \mathfrak{f}}^{\omega} (\vartheta)= \frac{i k R_0}{2 \pi}  \underset{\kappa+\rho~\textrm{even}}{\underset{\kappa \neq 0 } {\sum_{\kappa \rho}}} \mathscr{D}_{-\rho, \Delta M}^{\kappa \ast} (\varphi_{\varepsilon},\theta_{\varepsilon},0) \Xi_{\kappa 0} F_{\kappa \rho} J_{\vert \rho \vert} (k R_0 \vartheta)  \\
	\times \sum_{J J'} a_{J M}^{\tilde{J}} (\omega) b_{J' M'}^{\tilde{J'} \ast} (\omega) \sqrt{\frac{2J+1}{2J'+1}} C(J \kappa J' ; 0 0 0) C(J \kappa J' ; M \Delta M M'),
\end{multline}
where $\Delta M \equiv M' - M$ and $C(J_1,J_2,J_3;M_1,M_2,M_3)$ are Clebsch-Gordan coeffients \cite{Zare}. 
Since the atom-linear molecule potential is axially symmetric, only the $\Xi_{\kappa 0}$ coefficients contribute to the scattering amplitude. 

Eq.~(\ref{ScatAmplArbField}) can be simplified by limiting our considerations to special cases. A first such simplification arises when we let the initial free-rotor state to be the ground state, $\vert {J}, M \rangle \equiv \vert 0, 0 \rangle$. A second simplification is achieved by restricting the orientation of the electric field in the space-fixed frame to a particular geometry. 

(i) For an electric field \textit{parallel} to the initial wave vector, $\boldsymbol{\varepsilon} \upuparrows \mathbf{k}$, we have $\theta_{\varepsilon}\rightarrow 0$, $\varphi_{\varepsilon} \rightarrow 0$. As as result,
the Wigner matrix becomes $\mathscr{D}_{-\rho, \Delta M}^{\kappa \ast} (0,0,0)$,  which equals a Kronecker delta, $\delta_{-\rho \Delta M}$. Hence only the $\rho = - \Delta M'$ term yields a nonvanishing contribution to the scattering amplitude of Eq.~(\ref{ScatAmplArbField}),	

\begin{multline}
	\label{Ampl0bParallel}
	f_{0,0 \to \tilde{J'}, M' }^{\omega, \parallel}  (\vartheta) = J_{\vert M' \vert} (k R_0 \vartheta) \frac{i k R_0}{2 \pi}  \underset{\kappa+M'~\textrm{even}} {\sum_{\kappa \neq 0}}  \Xi_{\kappa 0} F_{\kappa M'} \\
	\times \sum_{J J'} a_{J 0}^{0} (\omega) b_{J' M'}^{\tilde{J'} \ast} (\omega) \sqrt{\frac{2J+1}{2J'+1}} C(J \kappa J' ; 0 0 0) C(J \kappa J' ; 0 M' M')
\end{multline}
We see that the angular dependence of the scattering amplitude for the parallel case is simple, given by a single Bessel function, $J_{\vert M' \vert}$.

(ii) For an electric field \textit{perpendicular} to the initial wave vector,  $\boldsymbol{\varepsilon} \perp \mathbf{k}$, we have $\theta_{\varepsilon}\rightarrow \frac{\pi}{2}$,~$\varphi_{\varepsilon} \rightarrow 0$. Hence
\begin{multline}
	\label{Ampl0bPerp}
	f_{0,0 \to \tilde{J'}, M' }^{\omega, \perp} (\vartheta) = \frac{i k R_0}{2 \pi}  \underset{\kappa+\rho~\textrm{even}}{\underset{\kappa \neq 0 } {\sum_{\kappa \rho}}}  d_{-\rho, M'}^{\kappa}\left( \frac{\pi}{2} \right) \Xi_{\kappa 0} F_{\kappa \rho} J_{\vert \rho \vert} (k R_0 \vartheta) \\ 
	\times \sum_{J J'} a_{J 0}^{0} (\omega) b_{J' M'}^{\tilde{J'} \ast} (\omega)\sqrt{\frac{2J+1}{2J'+1}} C(J \kappa J' ; 0 0 0)  C(J \kappa J' ; 0 M' M')
\end{multline}
where $ d_{-\rho, M'}^{\kappa}$ are the real Wigner rotation matrices. Since the summation mixes different Bessel functions (for a range of $\rho$'s), the angular dependence of the scattering amplitude in the perpendicular case is more involved than in the parallel case.

We note that, unfortunately, the Fraunhofer model does not distinguish between the parallel and antiparallel orientations of the field with respect to the initial wave vector, as can be seen by substituting $\mathscr{D}_{-\rho, \Delta M}^{\kappa \ast}(0, \pi, 0) = \delta_{\rho \Delta M} (-1)^{\kappa - \rho}$ into Eq.~(\ref{ScatAmplArbField}). This defect is inherent to the Fraunhofer model, since the diffraction occurs on a two-dimensional obstacle in the $XY$ plane, which looks the same from either side of the plane, no matter whether $\boldsymbol{\varepsilon} \upuparrows \mathbf{k}$ or $\boldsymbol{\varepsilon} \uparrow \downarrow \mathbf{k}$. 

\subsection{Results for $\text{Ne-OCS}(^1\Sigma, J=0 \to J')$ scattering in an electrostatic field}
\label{Ne-OCS}


We now proceed with the presentation of the collisional model with a concrete collision system in mind, namely He + OCS($^1\Sigma, J=0 \to J')$. The OCS molecule has been widely used in experiments with helium nanodroplets~\cite{Vilesov}.  The electric dipole moment $\mu = 0.709$ D, rotational constant $B = 0.2039$ cm$^{-1}$, and spectroscopic amicability make the OCS molecule a suitable candidate for an experiment to test the field-dependent Fraunhofer model for a $^1\Sigma$ molecule.

\begin{figure*}[htbp]
\centering\includegraphics[width=7cm]{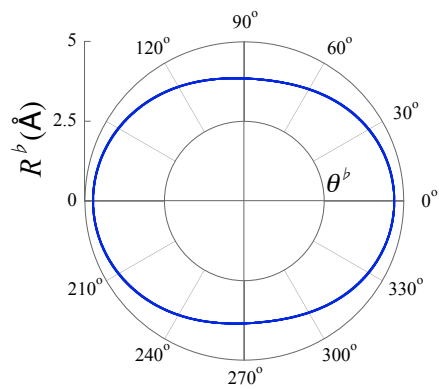}
\caption{Equipotential line $R^{\flat}(\theta^{\flat})$ for the Ne-OCS potential energy surface at a collision energy of 500 cm$^{-1}$. We note that the equipotential line for the Ar-NO collision system looks similar and is not shown. The Legendre moments for either potential energy surface are listed in Table \ref{table:legendre_coefs}.}\label{fig:PEScut}
\end{figure*}

According to Ref.~\cite{Zhu}, the ground-state Ne-OCS potential energy surface has a global minimum of a depth of  -81.26 cm$^{-1}$. In order to diminish the effect of this attractive well in the collision, we choose a collision energy $E_{\text{coll}}=500$ cm$^{-1}$, which corresponds to a wave number $k=21.09$ \r{A}$^{-1}$. The ``hard shell" of the potential energy surface at this collision energy, shown in Fig.~\ref{fig:PEScut}, we found by a fit to Eq.~(\ref{RhoExpBF}) for $\kappa \le 6$. The coefficients $\Xi_{\kappa 0}$ obtained from the fit are listed in the Table~\ref{table:legendre_coefs}. According to Eq.~(\ref{R0viab}), the $\Xi_{00}$ coefficient determines the hard-sphere radius $R_0$, which is responsible for elastic scattering.

\subsubsection{Differential cross sections}
\label{diffcrossOCS}
We start by analyzing the \textit{field-free} state-to-state differential cross section, which is given by
\begin{equation}
	\label{DiffCross00jmFF}
	\mathcal{I}_{0,0 \to J',M'}^{\text{f-f}}(\vartheta) =\vert f_{0,0 \to J',M'} (\vartheta) \vert^2 = \Phi_{J' \vert M' \vert}  \Xi_{J' 0}^2 J_{\vert M' \vert}^2 (k R_0 \vartheta)
\end{equation}
with
\begin{equation}
	\label{Philamnu}
	\Phi_{J' \vert M' \vert}= \left \{ 
	\begin{array}{ccl}
		\frac{(k R_0)^2}{4\pi}  \frac{\sqrt{(J'+\vert M' \vert)! (J'-\vert M' \vert)! }}{(J'+\vert M' \vert)!! (J'-\vert M' \vert)!! }  &    & \textrm{for $J'+\vert M' \vert$~even } \\ \\
						0 &    & \textrm{otherwise} 
	\end{array}    \right .
\end{equation}
see Eq. (\ref{InelAmplExpress}) and Ref.~\cite{Faubel}. We see that the state-to-state differential cross section is proportional to the square of the $\Xi_{J' 0}$ coefficient, which means that the shape of the repulsive potential provides a direct information about the relative probabilities of the field-free transitions and \textit{vice versa}. For the Ne-OCS system, the $\Xi_{2,0}$ coefficient dominates the anisotropic part of the potential, see Table~\ref{table:legendre_coefs}.  As a result, the corresponding $J=0 \to J'=2$ transition is expected to dominate the inelastic cross section. 

\begin{figure*}[htbp]
\centering
\includegraphics[width=8cm]{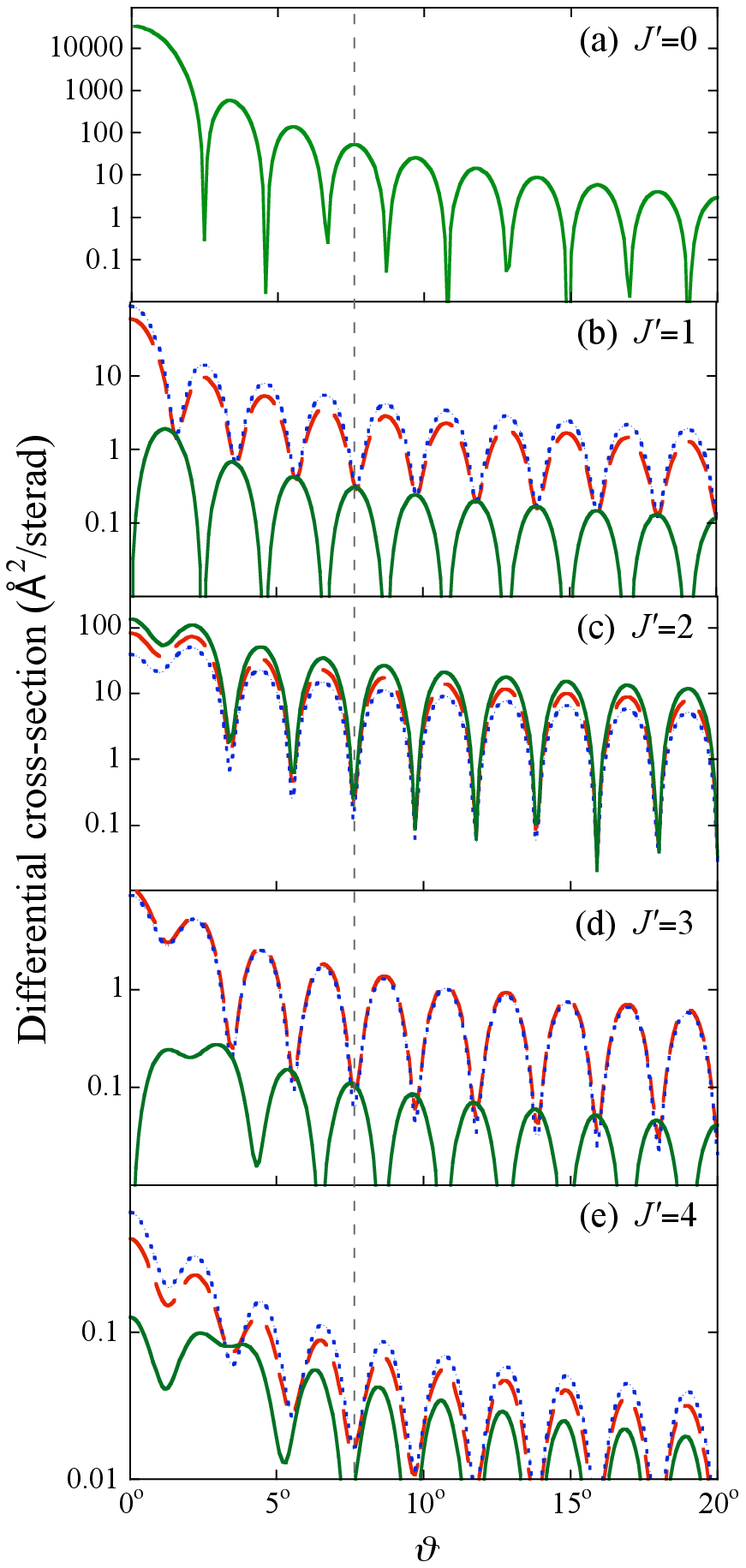}
\caption{Differential cross sections for the Ne-OCS $(J=0 \to J')$ collisions in an electrostatic field $\varepsilon$=50~kV/cm (red dashed line) and 100~kV/cm (blue dotted line), parallel to the relative velocity vector. The field-free cross sections are shown by the green solid line. The dashed vertical line serves to guide the eye in discerning the angular shifts of the partial cross sections.}
\label{fig:diff_parall}
\end{figure*}

\begin{figure*}[htbp]
\includegraphics[width=8cm]{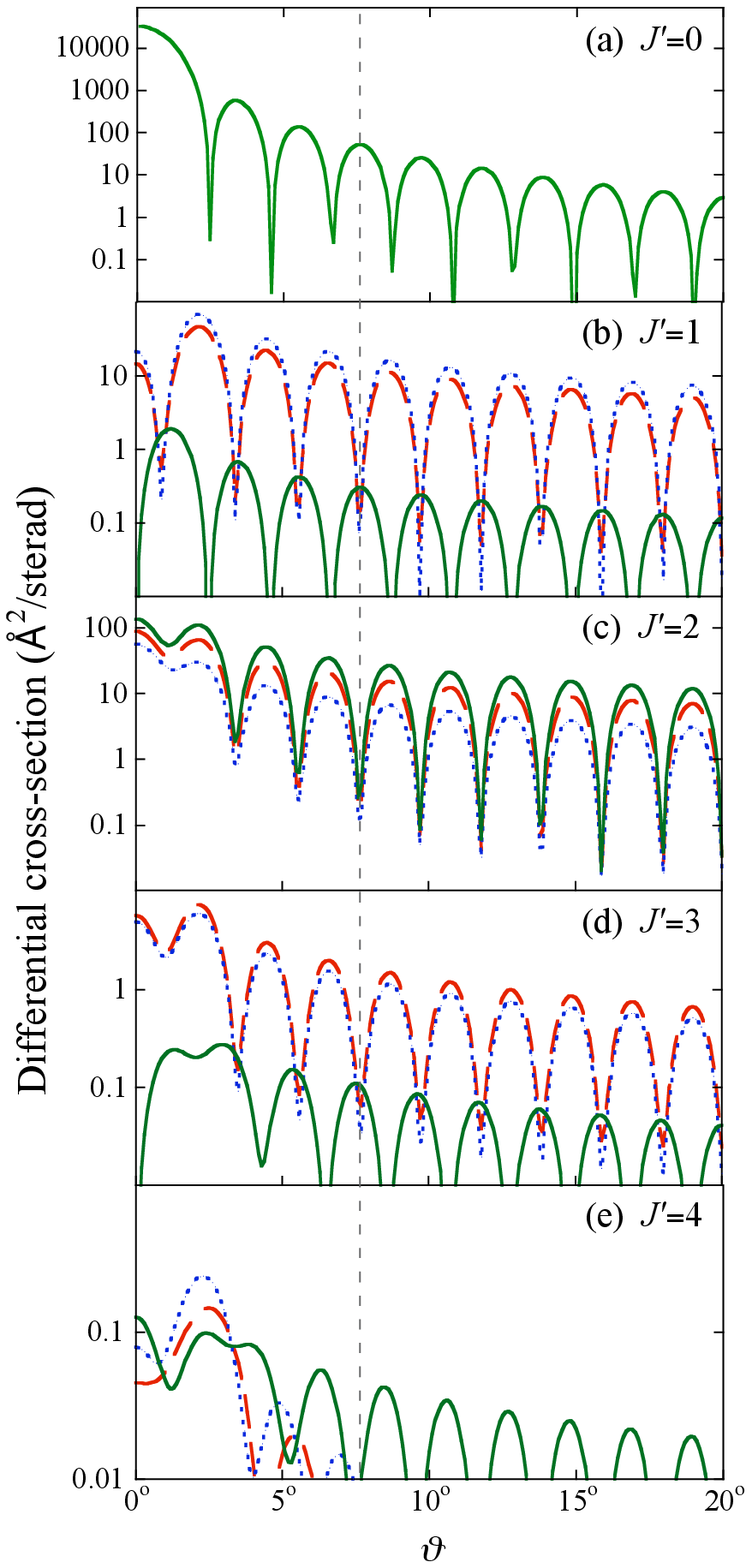}
\caption{Differential cross sections for the Ne-OCS $(J=0 \to J')$ collisions in an electrostatic field $\varepsilon$=50~kV/cm (red dashed line) and 100~kV/cm (blue dotted line), perpendicular to the relative velocity vector. The field-free cross sections are shown by the green solid line. The dashed vertical line serves to guide the eye in discerning the angular shifts of the partial cross sections.}
\label{fig:diff_perp}
\end{figure*}

Recalling the properties of the Bessel functions~\cite{Watson}, we see that for $kR_0\vartheta \gtrsim\frac{ \pi J'}{2}$ (which corresponds to $\vartheta \gtrsim J'$ degrees for the system under investigation), the differential cross-section has the following angular dependence:
\begin{equation}
	\label{AssympFiFrCrossJM}
	\mathcal{I}_{0,0 \to J',M'}^{\text{f-f}}(\vartheta)  \sim \left \{ 
	\begin{array}{ccl}
		\cos^2 \left(k R_0 \vartheta - \frac{\pi}{4} \right) &    & \textrm{ for $M'$~even } \\ \\
		\sin^2 \left(k R_0 \vartheta - \frac{\pi}{4}  \right) &    &\textrm{ for $M'$~odd} 
	\end{array}    \right .
\end{equation}
By averaging over $M'$ and taking into account that $\Phi_{J' \vert M' \vert}$ vanishes for $J'+\vert M' \vert$ odd, we obtain the angular dependence of the differential cross-section for a $0 \to J'$ transition:
 \begin{equation}
	\label{AssympFiFrCrossJaver}
	\mathcal{I}_{0 \to J'}^{\text{f-f}}(\vartheta)  \sim \left \{ 
	\begin{array}{ccl}
		\cos^2 \left(k R_0 \vartheta - \frac{\pi}{4}  \right) &    & \textrm{ for $J'$~even } \\ \\
		\sin^2 \left(k R_0 \vartheta - \frac{\pi}{4}  \right) &    &\textrm{ for $J'$~odd} 
	\end{array}    \right .
\end{equation}
The ``phase shift" of $\frac{\pi}{2}$ predicted by Eq. (\ref{AssympFiFrCrossJaver}) for the oscillations in the differential cross sections corresponding to even and odd field-free transitions is shown in Figs.~\ref{fig:diff_parall} and~\ref{fig:diff_perp}. The elastic scattering amplitude, given by Eq.~(\ref{AmplSphere}), has a $\sin^2 \left(k R_0 \vartheta - \frac{\pi}{4} \right)$ asymptote, and so is out of phase with even-$J'$-transitions. This latter effect, which is known as the ``Blair phase rule," can be also seen in Figs.~\ref{fig:diff_parall} and~\ref{fig:diff_perp}.
\\
\\The state-to-state differential cross sections for scattering in a field parallel ($\parallel$) and perpendicular ($\perp$) to $\mathbf{k}$ are given by
\begin{equation}
	\label{DiffCrossFieldsJaver}
	\mathcal{I}_{0 \to J'}^{\omega,(\parallel,\perp)}(\vartheta)=\sum_{M'} \mathcal{I}_{0,0 \to J',M'}^{\omega,(\parallel,\perp)}(\vartheta)
\end{equation}
with
\begin{equation}
	\label{DiffCrossFieldsJM}
	\mathcal{I}_{0,0 \to J',M'}^{\omega,(\parallel,\perp)}(\vartheta)=\left \vert f_{0,0 \to \tilde{J'}, M' }^{\omega, (\parallel,\perp)} (\vartheta)  \right \vert^2
\end{equation}
and are shown for the Ne+OCS collisions at $\varepsilon =50$ and $100$~kV/cm in Figs.~\ref{fig:diff_parall} and ~\ref{fig:diff_perp}. The figures reveal that an electrostatic field on the order of $10$~kV/cm dramatically alters the cross-sections. In this subsection we only analyze the field-induced ``phase shifts" of the oscillations, and defer the discussion of the amplitudes to subsection~\ref{sec:IntCrossSec} on the integral cross sections, to which the amplitudes are closely related.

(i) For a \textit{parallel field}, $\boldsymbol{\varepsilon} \parallel \mathbf{k}$, the differential cross section, Eq.~(\ref{DiffCrossFieldsJM}), has the same explicit angular dependence as for the field-free case, Eq.~(\ref{AssympFiFrCrossJM}). However, the field suppresses the selection rule~(\ref{Philamnu}) and so the summation in Eq.~(\ref{DiffCrossFieldsJaver}) comprises all $M'$-states. Therefore, the resulting cross section is a field-dependent mixture of the sine- and cosine-contributions given by Eq.~(\ref{AssympFiFrCrossJM}). 
The angular dependence of the differential cross sections in Fig.~\ref{fig:diff_parall} can be gleaned from Eq.~(\ref{Ampl0bParallel}). The first sum in Eq.~(\ref{Ampl0bParallel}) extends over even $\kappa$ for even $M'$, and over odd $\kappa$ for odd $M'$. Therefore, the $\Xi_{\kappa 0}$ coefficients, Table~\ref{table:legendre_coefs}, determine not only the  relative contributions of different $J'$ states, but also of different $M'$ states in Eq.~(\ref{DiffCrossFieldsJaver}). Since the $\Xi_{2 0}$ coefficient eclipses the others, transitions to even $M'$ states dominate whenever the field is high enough, and the field-free cross-section~(\ref{AssympFiFrCrossJaver}) has a $\cos^2\vartheta$ asymptote. This can be clearly seen in Fig.~\ref{fig:diff_parall}: for odd $J'$, there is a field-induced phase shift of the differential cross section, which is absent for transitions to even $J'$.

(ii) For a \textit{perpendicular field}, $\boldsymbol{\varepsilon} \perp \mathbf{k}$, several Bessel functions contribute to the scattering amplitude. However, since the summation in Eq.~(\ref{Ampl0bPerp}) requires that $\kappa+\rho$ be even, it is the even Bessel functions which, like in the case of a parallel field, can be expected to dominate the cross section. Indeed, the cross sections shown in Figs.~\ref{fig:diff_parall} and~\ref{fig:diff_perp} for parallel and perpendicular fields are, for  $J'=1,2,3$, similar to one another. However, the $J=0 \to J'=4$ differential cross section in the perpendicular field exhibits an additional phase shift. This cross section represents a special case as it is not dominated by the $\Xi_{2 0}$ coefficient. The $\Xi_{2 0}$ coefficient fails to dominate the $J=0 \to J'=4$ cross section because of the selection rule, $J'=J; J \pm 2$, that the Clebsch-Gordan coefficients $C(J 2 J', 0 0 0)$ impose on the $\kappa=2$ term. However, the products of the hybridization coefficients, $a_{J 0}^{0} (\omega) b_{J 0}^{\tilde{4}\ast} (\omega)$ and $a_{J 0}^{0} (\omega) b_{J\pm2, 0}^{\tilde{4}\ast} (\omega)$ that occur in the term are very small, due to a tiny overlap of the $a_{J 0}^{0} (\omega)$ and $ b_{J' 0}^{\tilde{4}\ast} (\omega)$ hybridization coefficents. Therefore, a superposition of both even and odd Bessel functions contributes to the  $J=0 \to J'=4$ differential cross section. A more detailed discussion of the overlaps of the hybridization coefficients follows in the next subsection. 


\subsubsection{Integral cross sections}
\label{sec:IntCrossSec}

\begin{figure*}[htbp]
\includegraphics[width=15cm]{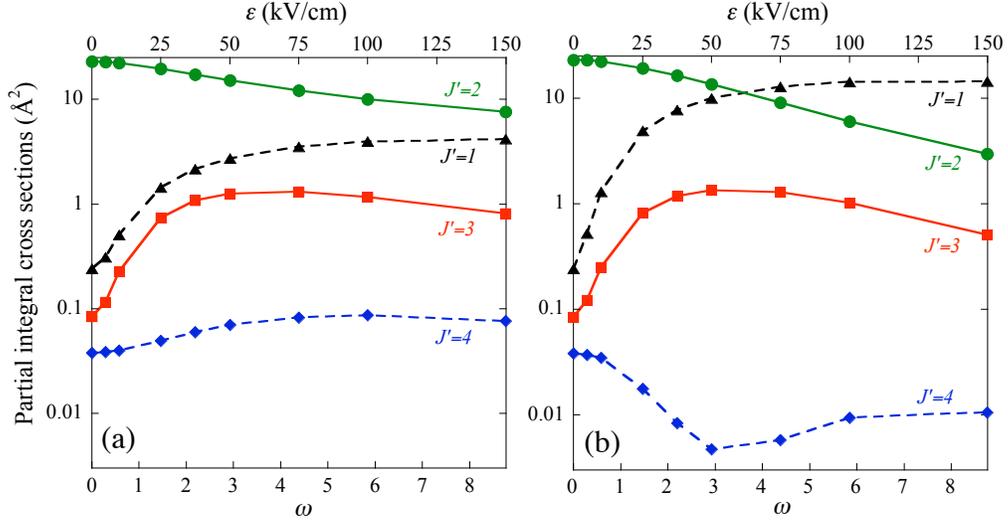}
\caption{Partial integral cross sections for Ne-OCS ($J=0 \to J'$) collisions in an electric field parallel, panel~(a), and perpendicular, panel~(b), to the initial wave vector.}\label{fig:integral_cross}
\end{figure*}

\begin{figure*}[htbp]
\centering\includegraphics[width=8cm]{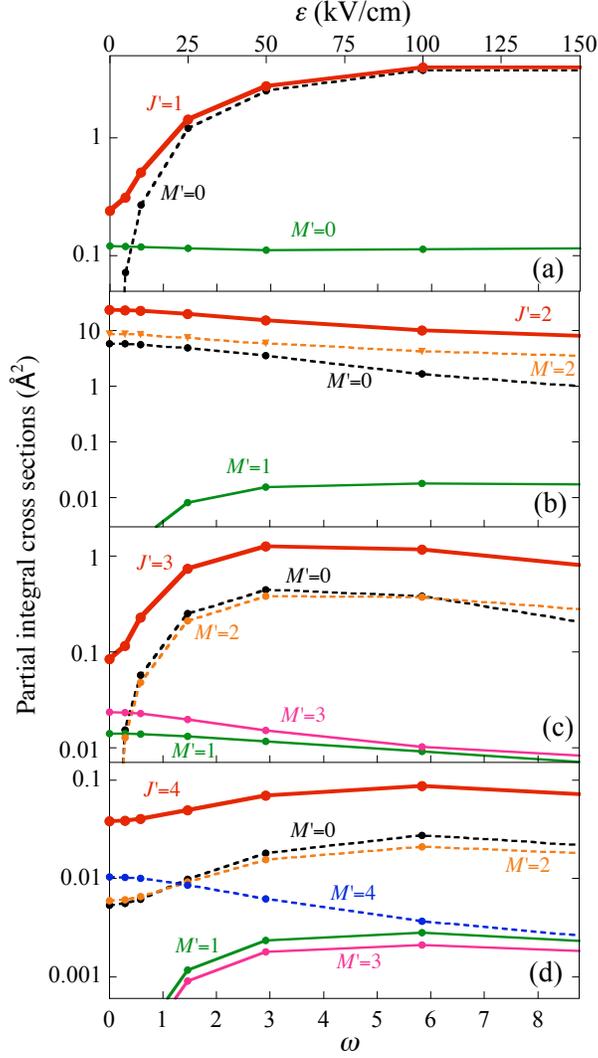}
\caption{Partial integral cross sections for Ne-OCS ($J=0, M=0 \to J', M'$) collisions in an electric field parallel to the initial wave vector. The red solid lines show the $M'$-averaged cross sections for the $J=0 \to J'$ collisions.}\label{fig:partial_jm}
\end{figure*}

The angular range, $\vartheta \lesssim 30^{\circ}$, where the Fraunhofer approximation applies the best, comprises the largest impact-parameter collisions that contribute to the scattering the most, see Figs.~\ref{fig:diff_parall} and~\ref{fig:diff_perp}. Therefore, the integral cross section can be obtained to a good approximation by integrating the Fraunhofer differential cross section, Eq. (\ref{DiffCrossFieldsJaver}), over the solid angle $\sin\vartheta d \vartheta d \varphi$,
\begin{equation}
	\label{IntCrossSec}
	\sigma_{0 \to J'}^{\omega,(\parallel,\perp)} = \int_{0}^{2\pi} d\varphi \int_{0}^{\pi} \mathcal{I}_{0 \to J'}^{\omega,(\parallel,\perp)}(\vartheta) \sin\vartheta d \vartheta
\end{equation}
The integral cross-sections thus obtained for the field parallel and perpendicular to the initial wave vector are presented in Fig.~\ref{fig:integral_cross}.  One can see that the state-to-state cross section for the  $J=0 \to J'=2$ collisions steadily decreases with the interaction parameter $\omega$, whereas the other state-to-state cross sections show a non-monotonous dependence. These features reflect the dependence of the integral cross sections on the final $M'$ states accessed by the inelastic collisions, which is shown in Fig.~\ref{fig:partial_jm} for the $\boldsymbol{\varepsilon} \parallel \mathbf{k}$ case. 
The relative weights of different $M'$ states contributing to the cross sections are determined by a combination of the $\Xi_{\kappa 0}$ coefficients of Eq.~(\ref{InelAmplExpress}) and the $F_{\kappa, M'}$ factor of Eq.~(\ref{Flamnu}). Since  the $\Xi_{20}$ coefficient looms over the rest and the $F_{\kappa, M'}$ factor is only non-vanishing for $\kappa+M'$ even and $\kappa \ge M'$, collisions leading to $M'=0,2$ dominate for all $J'$ values, see Fig.~\ref{fig:partial_jm}. 

\begin{figure*}[htbp]
\centering\includegraphics[clip,width=7cm]{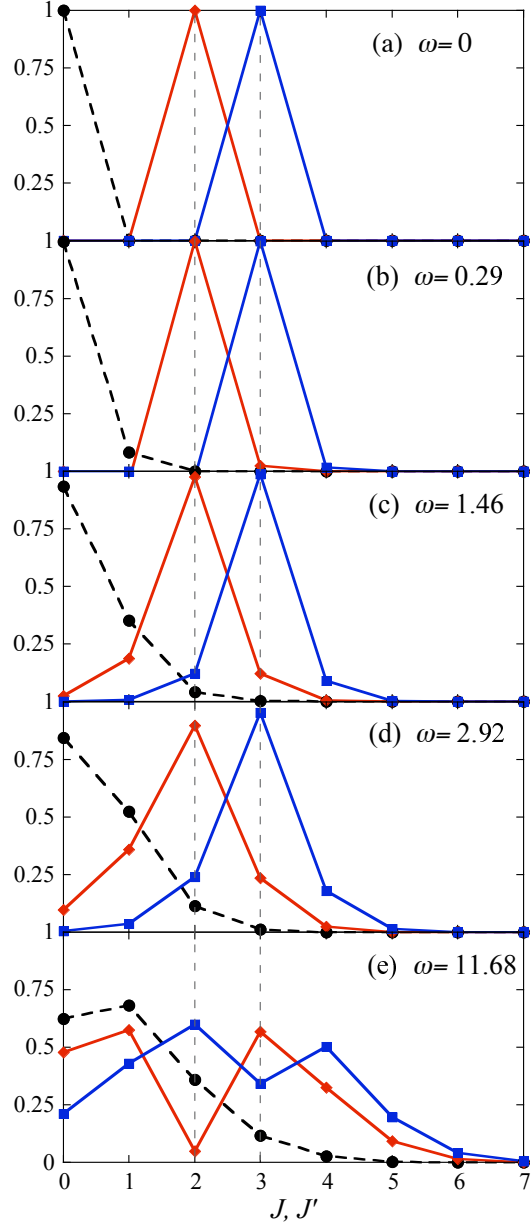}
\caption{Absolute values of the hybridization coefficients $a_{J 0}^{0} (\omega)$ (black dashed line), $b_{J' 0}^{\tilde{2}}$ (red solid line) and $b_{J' 0}^{\tilde{3}}$ (blue solid line) for different values of the interaction parameter $\omega$. The dashed vertical lines serves as a guide to the eye. See text.}\label{fig:aj_coefs}
\end{figure*}

But how to explain the field dependence of the cross sections for the prevalent $J=0, M=0 \to J',M'=0,2$ collisions? The answer comes from the realization that the dependence of the integral cross sections on $\omega$ is entirely determined by the coefficients $a_{J M}^{\tilde{J}} (\omega)$ and $b_{J' M'}^{\tilde{J'}}$ in the scattering amplitude -- both for $\boldsymbol{\varepsilon} \parallel \mathbf{k}$, Eq.~(\ref{Ampl0bParallel}), and $\boldsymbol{\varepsilon} \perp \mathbf{k}$, Eq. (\ref{Ampl0bPerp}). What the field dependence of the coefficients $a_{J 0}^{0} (\omega)$, $b_{J' 0}^{\tilde{2}} (\omega)$, and $b_{J' 0}^{\tilde{3}} (\omega)$ looks like is shown in~Fig.~\ref{fig:aj_coefs}. As noted in subsection~\ref{diffcrossOCS}, for $J'=1,2,3$ the scattering amplitude of Eq.~(\ref{Ampl0bParallel}) is dominated by the $\Xi_{20}$ coefficient, which entails the selection rule $J'=J; J \pm 2$ in the summation over $J$ and $J'$. In the field-free case, this selection rule is only satisfied for the $J=0 \to J'=2$ scattering, which indeed governs the field-free collisions. Once the field is applied, the ``distributions" of the $a_{J 0}^{0} (\omega)$ and $b_{J 0}^{\tilde{J'}} (\omega)$  coefficients undergo a broadening, see Fig.~\ref{fig:aj_coefs}~(b)-(e). For the $J'=2$ channel, such a broadening reduces the overlap of the corresponding hybridization coefficients and thus diminishes the $J=0 \to J'=2$ cross section. In contradistinction, the overlap of the hybridization coefficients for $J'=1,3$ increases with $\omega$, resulting in an increase of $J=0 \to J'=1,3$ cross sections, see Fig.~\ref{fig:integral_cross}. At even higher $\omega$, the spread of the coefficients is so large that the products $a_{J 0}^{0} (\omega) b_{J' 0}^{\tilde{J'} \ast} (\omega)$, corresponding to the selection rule $J'=J; J \pm 2$, become very small, cf. Fig.~\ref{fig:aj_coefs}(e). As a result, the cross-sections for the $J'=1,3$ channels decrease. The field dependence of the $J=0 \to J'=4$ channel is less straightforward, since, as outlined above, its cross section is governed by $\Xi_{\kappa 0}$ coefficients with $\kappa \ne 2$.

A prominent feature in Fig.~\ref{fig:integral_cross} is the significant influence of the orientation of $\boldsymbol{\varepsilon}$ with respect to $\mathbf{k}$ on the cross section for the $J=0 \to J'=1$ channel. As mentioned above, for both $\boldsymbol{\varepsilon} \parallel \mathbf{k}$ and $\boldsymbol{\varepsilon} \perp \mathbf{k}$, the partial $J=0;M=0 \to J'=1; M'=0$ cross section provides the main contribution to the $J=0 \to J'=1$ transition. However, an inspection of Eqs.~(\ref{Ampl0bParallel}) and~(\ref{Ampl0bPerp}) reveals that the integral cross sections for the $J=0;M=0 \to J'=1;M'=0$ transitions is always greater for $\boldsymbol{\varepsilon} \perp \mathbf{k}$ than for $\boldsymbol{\varepsilon} \parallel \mathbf{k}$, due to the coefficients $d^{\kappa}_{-\rho, 0}(\frac{\pi}{2})$.

\subsubsection{Frontal versus lateral steric asymmetry}
\begin{figure*}[htbp]
\centering\includegraphics[width=7.5cm]{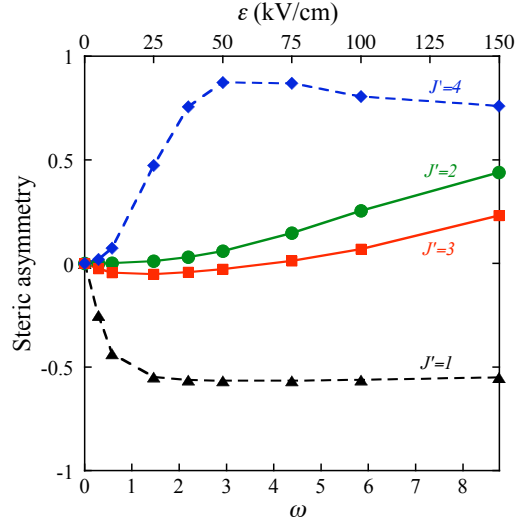}
\caption{Steric asymmetry, as defined by Eq.~(\ref{StericAsymmetry}), for Ne-OCS~($J=0 \to J'$) collisions.}\label{fig:asymmetry}
\end{figure*}

We define the steric asymmetry as
\begin{equation}
	\label{StericAsymmetry}
 	S_{\mathfrak{i} \to \mathfrak{f}} = \frac{\sigma_{\parallel}-\sigma_{\perp}}{\sigma_{\parallel}+\sigma_{\perp}},
\end{equation}
where the cross sections $\sigma_{\parallel, \perp}$ correspond, respectively, to $\boldsymbol{\varepsilon} \parallel \mathbf{k}$ and $\boldsymbol{\varepsilon} \perp \mathbf{k}$. The dependence of the steric asymmetry on the permanent dipole interaction parameter $\omega$  is presented in Fig.~\ref {fig:asymmetry}. One can see that a particularly pronounced asymmetry obtains for the $J=0 \to J'=1,4$ collisions. This can be traced to the field dependence of the corresponding integral cross sections, Fig.~\ref{fig:integral_cross}. Within the Fraunhofer model, elastic collisions do not exhibit any steric asymmetry. This follows from the isotropy of the elastic scattering amplitude, Eq. (\ref{Ampl0bPerp}), which depends on the radius $R_0$ only: a sphere looks the same from any direction.

\section{The Fraunhofer model of rotationally inelastic scattering of symmetric-top-equivalent linear polar molecules by closed-shell atoms in an electrostatic field}
\label{sec:Fraun2Pi}

\subsection{The field-dependent scattering amplitude}
\label{sec:scatamplsymtop}

Here we consider a symmetric top-equivalent linear polar molecule, such as $\text{OH}(^2 \Pi_{\frac{1}{2}})$, $\text{NO}(^2 \Pi_{\frac{1}{2}})$, colliding with a closed-shell atom.  We treat the molecule as a pure Hund's case~(a) species, characterized by a non-zero projection, $\Omega$, of the electronic angular momentum on the molecular axis, whose definite-parity rotational wavefunction is given by
\begin{equation}
	\label{SymmetrWF}
	\vert J, M, \vert \Omega \vert, \epsilon \rangle = \frac{1}{\sqrt{2}} \biggl [  \vert J, M,  \vert \Omega \vert  \rangle  + \epsilon \vert J, M,  - \vert \Omega \vert  \rangle \biggr ]
\end{equation}
where the symmetry index $\epsilon$ distinguishes between the members of a given $\Omega$ doublet. The symmetry index takes the value of $+1$  for $e$ levels and of $-1$ for $f$ levels. The parity of the wave function is equal to $\epsilon (-1)^{J-\frac{1}{2}}$~\cite{Brown75}.

\begin{figure*}[htbp]
\centering\includegraphics[width=15.5cm]{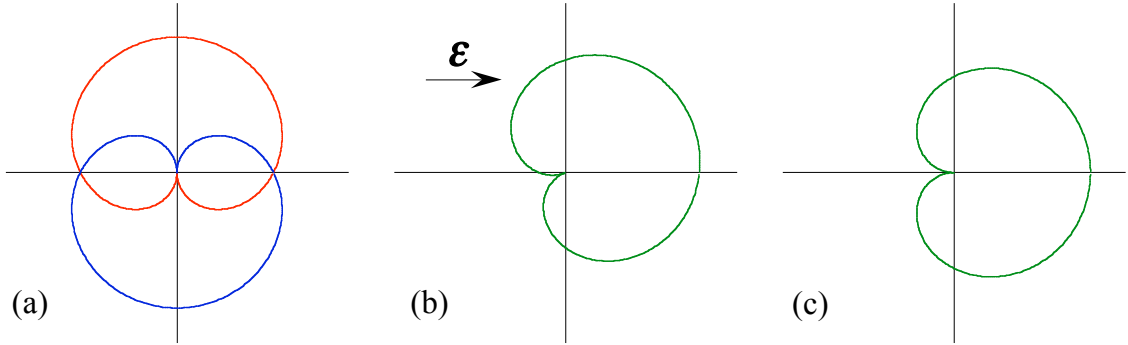}
\caption{The moduli of the symmetric-top wavefunction for $J=M=\Omega=\frac{1}{2}$. Panel (a) shows the field-free wavefunctions, Eq.~(\ref{SymmetrWF}), for
$\epsilon=1$ (blue line) and $\epsilon=-1$ (red line). Panels (b) and (c) show the wavefunctions of the precessing states in the field, Eq.~(\ref{OrientedWF}),
for, respectively, an incomplete ($\alpha=0.832$, $\beta = 0.555$) and perfect ($\alpha=\beta =\frac{1}{\sqrt{2}}$)
mixing of the $\Omega$ doublet states. See text.}
\label{fig:precess}
\end{figure*}

The rotational states of a Hund's case (a) molecule with $J>0$ and $M>0$ can be oriented by coupling the opposite-parity members of an $\Omega$ doublet via the electric-dipole interaction.  Such a coupling creates \textit{precessing states}, in which the body-fixed electric dipole moment $\boldsymbol{\mu}$ precesses about the field vector. As a result, molecular rotation does not average out the dipole moment in first order. A precessing state
is a hybrid of the two opposite-parity members of an $\Omega$-doublet, and can be written as
\begin{equation}
	\label{OrientedWF}
	\vert J, M, \vert \Omega \vert, w \rangle = \alpha (w) \vert J, M, \vert \Omega \vert, \epsilon=-1 \rangle +  \beta(w) \vert J, M, \vert \Omega \vert, \epsilon=1 \rangle,
\end{equation}
with $w \equiv \mu \varepsilon /\Delta$ an interaction parameter which measures the maximum potential energy of the electric dipole in terms of the $\Omega$-doublet splitting, $\Delta$, for $J=\vert \Omega \vert$. For a precessing state with $w \gg 1$, the coefficients $|\alpha(w)| =  |\beta(w)| = 2^{-\frac{1}{2}}$, and the mixing of the states within an $\Omega$ doublet is perfect. A less perfect mixing,  $ |\alpha(w)| \neq |\beta(w)|$, obtains when $w \le 1$. The wavefunction, Eq. (\ref{OrientedWF}), reduces for a precessing state with a perfect mixing to $\vert J, M, \vert \Omega \vert \rangle$. It is the inherent orientation of the precessing states along with their mixed parity that enters the Fraunhofer model for the scattering of Hund's case (a) molecules in an electric field. The directionality of the precessing states is illustrated in Figure~\ref{fig:precess}. We assume the hybridization of $J$-states for a symmetric-top state to be negligible.

 The symmetric top wavefunction is given by a Wigner rotation matrix
 \begin{equation}
	\label{RotatWF}
	\vert J, M, \Omega \rangle = \sqrt{\frac{2 J+1}{4 \pi}}  \mathscr{D}_{M \Omega}^{J \ast} (\varphi^{\sharp}, \theta^{\sharp}, \gamma^{\sharp}=0)
\end{equation}
In analogy with Eqs.~(\ref{PendularStateASpFix}) and (\ref{PendularStateBSpFix}), we transform the wave function to the space-fixed  frame,
\begin{equation}
	\label{FieldToSpaceDmatrix}
	\mathscr{D}_{M \Omega}^{J \ast} (\varphi^{\sharp}, \theta^{\sharp}, 0) = \sum_{\xi} \mathscr{D}_{\xi M}^{J} (\varphi_{\varepsilon}, \theta_{\varepsilon}, 0) \mathscr{D}_{\xi \Omega}^{J \ast} (\varphi, \theta, 0) 
\end{equation}
For transitions with $| \Omega |=|\Omega'|$, the initial and final precessing states can, therefore, be written as
\begin{multline}
	\label{PrecessingStateASpFix}
	\vert \mathfrak{i} \rangle = \sqrt{\frac{2J+1}{4 \pi}} \sum_{\xi} \mathscr{D}_{\xi M}^{J} (\varphi_{\varepsilon},\theta_{\varepsilon},0) \frac{1}{\sqrt{2}} \Biggl \{  \left [ \alpha(w)+ \beta(w)  \right ] \mathscr{D}_{\xi |\Omega|}^{J \ast} (\varphi, \theta, 0) \\
	+ \left [ - \alpha(w)+ \beta(w) \right ] \mathscr{D}_{\xi -|\Omega|}^{J \ast} (\varphi, \theta, 0) \Biggr \}
\end{multline}

\begin{multline}
	\label{PrecessingStateBSpFix}
	\langle \mathfrak{f} \vert = \sqrt{\frac{2J'+1}{4 \pi}} \sum_{\xi'} \mathscr{D}_{\xi' M'}^{J' \ast} (\varphi_{\varepsilon},\theta_{\varepsilon},0) \frac{1}{\sqrt{2}} \Biggl \{  \left [ \alpha'(w)+\beta'(w)  \right ] \mathscr{D}_{\xi' |\Omega|}^{J'} (\varphi, \theta, 0) \\
	+ \left [ - \alpha'(w)+\beta'(w)  \right ] \mathscr{D}_{\xi' -|\Omega|}^{J'} (\varphi, \theta, 0) \Biggr \}	
\end{multline}

By substituting from Eqs.~(\ref{PrecessingStateASpFix}) and~(\ref{PrecessingStateBSpFix}) into Eq.~(\ref{InelAmplExpress}), we finally obtain the scattering amplitude for inelastic collisions of symmetric-top molecules in precessing states
\begin{multline}
	\label{ScatAmplLamDoubl}	
	f_{\mathfrak{i} \to \mathfrak{f}}^{w} (\vartheta)= \frac{i k R_0}{4 \pi} \sqrt{\frac{2J+1}{2J'+1}}  \underset{\kappa+\rho~\textrm{even}}{\underset{\kappa \neq 0 } {\sum_{\kappa \rho}}}  \mathscr{D}_{-\rho, \Delta M}^{\kappa \ast}  (\varphi_{\varepsilon},\theta_{\varepsilon},0) \Xi_{\kappa 0} F_{\kappa \rho} J_{\vert \rho \vert} (k R_0 \vartheta) \\
	\times C(J \kappa J' ; |\Omega| 0 |\Omega|)  C(J \kappa J' ; M \Delta M M')  \Biggl \{ \left[\alpha(w) \alpha'(w) +\beta(w) \beta'(w)   \right] \biggl[ (-1)^{\kappa} + (-1)^{\Delta J} \biggr] \\ 
	+ \left[ \alpha(w) \beta'(w)  +  \alpha'(w) \beta(w)   \right] \biggl[ (-1)^{\kappa} - (-1)^{\Delta J} \biggr]   \Biggr \}
\end{multline}
We note that if both the initial and final precessing states are perfectly mixed, the term in the curly brackets of Eq. (\ref{ScatAmplLamDoubl}) reduces to $2(-1)^\kappa$. The scattering amplitudes for different orientations of the electrostatic field $\boldsymbol{\varepsilon}$ with respect to the initial wave vector $\mathbf{k}$ are obtained from Eq.~(\ref{ScatAmplLamDoubl}) by substituting the appropriate values of the angles: $\theta_{\varepsilon}=0 ; \varphi_{\varepsilon}=0$ for $\boldsymbol{\varepsilon} \parallel \mathbf{k}$, and $\theta_{\varepsilon}=\frac{\pi}{2} ; \varphi_{\varepsilon}=0$ for $\boldsymbol{\varepsilon} \perp \mathbf{k}$. Eq.~(\ref{ScatAmplLamDoubl}) implies that the integral cross-sections, cf. Eqs. (\ref{DiffCrossFieldsJaver}) - (\ref{IntCrossSec}),  for  $J \to J'$ transitions are the same in the parallel and perpendicular fields. However, the partial integral cross sections for $J,M \to J',M'$ transitions do depend on whether the field is parallel or perpendicular to $\mathbf{k}$.

\subsection{Results for $\text{Ar-NO} (^2\Pi_{\frac{1}{2}})$ scattering in an electrostatic field}
\label{sec:Ar-NO}

We now consider the excitation of $\text{NO}(J=|\Omega|=\frac{1}{2}, f \to J', |\Omega|, e/f)$ by collisions with Ar, under conditions similar to those defined in Refs.~\cite{vanLeuken96} and~\cite{Alexander93}: a hexapole state selector selects the $\epsilon=-1 (f)$ state, Eq.~(\ref{SymmetrWF}), which adiabatically evolves into a partially oriented state, Eq.~(\ref{OrientedWF}), when the collision system enters the electric field. The electric field of 16 kV/cm, directed parallel to the initial wave vector, $\boldsymbol{\varepsilon} \parallel \mathbf{k}$, creates a precessing state, Eq.~(\ref{OrientedWF}), whose hybridization coefficients are $\alpha = 0.832$ and $\beta = 0.555$. Next, a collision with an Ar atom excites the NO molecule to a final, field-free state, $\vert J',M',|\Omega|, \epsilon' \rangle$. 
The final, excited state is considered to be exempt from any effects of the electric field, as its $\Omega$-doubling is large and hence $w<1$. As a result, $\beta'(w)=1$ or $\alpha'(w)=1$ for a final state of $e$ or $f$ parity, respectively. For a collision so defined, the scattering amplitude, Eq.~(\ref{ScatAmplLamDoubl}) for $\boldsymbol{\varepsilon} \parallel \mathbf{k}$, takes the form
\begin{multline}
	\label{ScatAmplNOAr}	
	f_{\mathfrak{i} \to \mathfrak{f}}^{w,\parallel} (\vartheta)= \frac{i k R_0}{4 \pi} \sqrt{\frac{2}{2J'+1}}  J_{\vert \Delta M \vert} (k R_0 \vartheta) \underset{\kappa+\Delta M~\textrm{even}} {\sum_{\kappa \neq 0 }}  \Xi_{\kappa 0} F_{\kappa, \Delta M} C(J \kappa J' ; M \Delta M M') C(J \kappa J' ; |\Omega| 0 |\Omega|) \\
	\times \left \{  \begin{array}{c}	
	 \beta(w)  \biggl[ (-1)^{\kappa} + (-1)^{\Delta J} \biggr] + \alpha(w) \biggl[ (-1)^{\kappa} - (-1)^{\Delta J} \biggr] \\ \\
	 \alpha(w) \biggl[ (-1)^{\kappa} + (-1)^{\Delta J} \biggr] + \beta(w) \biggl[ (-1)^{\kappa} - (-1)^{\Delta J} \biggr]
	 \end{array}  \right \}
\end{multline}
where the first or second row of the expression in the curly brackets corresponds to a final state of, respectively, $e$ or $f$ parity. The coefficients $\Xi_{\kappa 0}$ of the Ar-NO interaction potential, extracted from the data of Sumiyoshi~\textit{et al.}~\cite{Sumiyoshi07}, are listed in Table~\ref {table:legendre_coefs}. According to ref.~\cite{Sumiyoshi07}, the Ar-NO potential surface exhibits a global minimum of $-115.4$ cm$^{-1}$ and, thus, the Fraunhofer model should be valid at collision energies $E_{\text{coll}}>400$ cm$^{-1}$. In ref.~\cite{Stolte}, the rigid shell QQT model was also used at these energies.

\begin{figure*}[htbp]
\includegraphics[width=15cm]{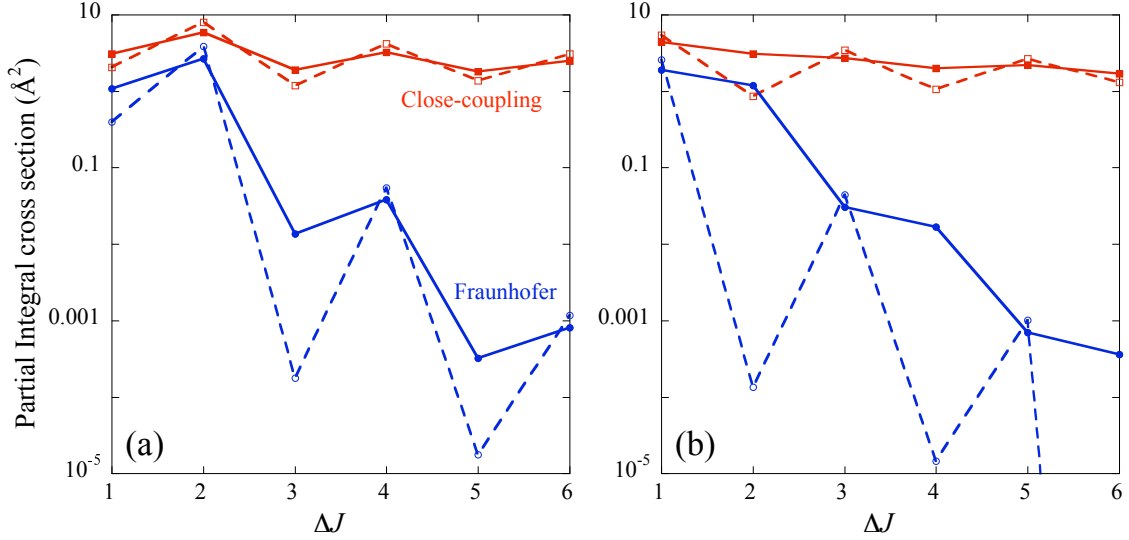}
\caption{Integral cross sections for the excitation of NO($J=|\Omega|=\frac{1}{2},f$) in collisions with Ar to higher rotational levels of the $|\Omega|=\frac{1}{2}$ manifold. Panels (a) and (b) pertain, respectively, to parity-conserving and parity-breaking Ar-NO collisions.  The results obtained from the Fraunhofer model are shown by blue curves, those obtained from the close-coupling calculations of Ref.~\cite{Alexander93} by red curves. Dashed lines pertain to field-free scattering, solid lines to scattering in an electric field $\varepsilon =16$ kV/cm.}
\label{fig:NO-Ar_cross}
\end{figure*}

The state-to-state integral cross sections for spin-conserving collisions ($| \Omega' |=| \Omega |=\frac{1}{2}$) at a collision energy of 442 cm$^{-1}$ are shown in Fig.~\ref{fig:NO-Ar_cross}, along with the close coupling calculations of Refs.~\cite{vanLeuken96} and~\cite{Alexander93}. The analytic Fraunhofer model provides a simple interpretation of the features exhibited by the cross sections.

First, let us consider the field-free case for an initial $f$ state, i.e., for $\alpha(w=0) =1$ and $\beta(w=0) = 0$:
\begin{multline}
	\label{ScatAmplNOArFF}	
	f_{\mathfrak{i} \to \mathfrak{f}}^{w=0} (\vartheta) \sim  \\
	J_{\vert \Delta M \vert} (k R_0 \vartheta) \underset{\kappa+\Delta M~\textrm{even}} {\sum_{\kappa \neq 0 }}  \Xi_{\kappa 0} F_{\kappa, \Delta M} C(J \kappa J' ; M \Delta M M') C(J \kappa J' ; |\Omega| 0 |\Omega|)
	 \left \{  \begin{array}{c}	
	  \biggl[ (-1)^{\kappa} - (-1)^{\Delta J} \biggr] \\ \\
	 \biggl[ (-1)^{\kappa} + (-1)^{\Delta J} \biggr] 
	 \end{array}  \right \}
\end{multline}
Eq. (\ref{ScatAmplNOArFF}) immediately reveals that if the potential energy surface is governed by terms with $\kappa$ even, parity-conserving transitions, $f \to f$, will dominate for $\Delta J$ even, while parity-changing transitions, $f \to e$, will dominate for $\Delta J$ odd. This propensity can be seen in Fig.~\ref{fig:NO-Ar_cross}. It was explained previously in Ref.~\cite{Esposti95} by a rather involved analysis of the close-coupling matrix elements.

The qualitative features of the scattering in an electric field can also be readily explained by the Fraunhofer model. If the field is present and the target molecule oriented, both even and odd $\Delta J$'s in the curly brackets of~Eq.~(\ref{ScatAmplNOAr}) contribute to the scattering for any value of $\kappa$. For a potential energy surface governed by even-$\kappa$ terms, the electric field will enhance the parity-conserving transitions for $\Delta J$ odd, and suppress them for $\Delta J$ even; parity-breaking collisions will prevail for $\Delta J$ even and subside for $\Delta J$ odd.

From Fig.~\ref{fig:NO-Ar_cross} one could see that for $\Delta J >2$ the Fraunhofer model yields an integral cross section which is significantly smaller than the one obtained from a close-coupling calculation. This is supported by the work of Aoiz \textit{et al.}~\cite{Aoiz03}, who found that the diffractive contribution to the differential cross sections of Ar-NO collisions is much greater for $\Delta J = 2$ than for $\Delta J = 3$.

As noted in subsection~\ref{sec:scatamplsymtop}, the integral cross-sections for $J \to J'$ scattering are the same for $\boldsymbol{\varepsilon} \parallel \mathbf{k}$ and $\boldsymbol{\varepsilon} \perp \mathbf{k}$. Therefore, the Fraunhofer model does not distinguish  between the two collisional configurations and yields a zero steric asymmetry as defined by Eq. (\ref{StericAsymmetry}) for symmetric-top-equivalent molecules.

\begin{figure*}
\includegraphics[width=7cm]{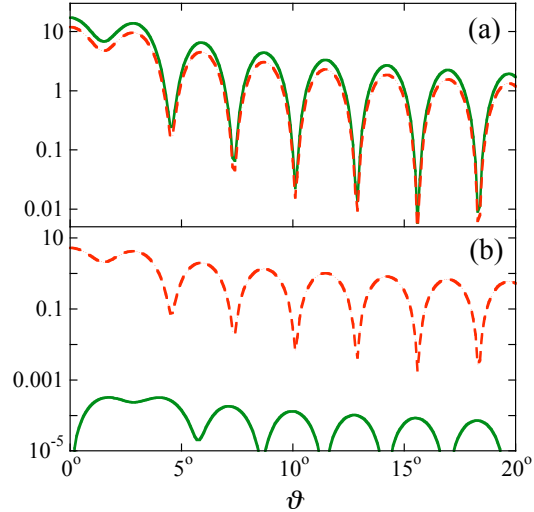}
\caption{Differential cross sections for the excitation of NO($J=|\Omega|=\frac{1}{2},f$) in collisions with Ar to the $J'=\frac{5}{2}, \Omega=\frac{1}{2}$ state. Panels (a) and (b) pertain, respectively, to parity-conserving and parity-breaking Ar-NO collisions.  
Green lines pertain to field-free scattering, red lines to scattering in an electric field $\varepsilon =16$~kV/cm.}
\label{fig:NO-Ar_diff}
\end{figure*}

Eqs.~(\ref{ScatAmplNOAr}) and~(\ref{ScatAmplNOArFF}) also reveal the angular dependence of the scattering. In particular, by making use of the asymptotic forms of the Bessel functions~\cite{Watson}, we see that for a potential energy surface governed either by even- or odd-$\kappa$ terms, the differential cross sections for parity-conserving and parity-breaking transitions will be out of phase. This is illustrated for scattering from $\vert J=\frac{1}{2},|\Omega|=\frac{1}{2}, f \rangle$ to $\vert J'=\frac{5}{2}, |\Omega|=\frac{1}{2}, e/f \rangle$ states in Fig.~\ref{fig:NO-Ar_diff} (full curves). We also note that the parity-breaking cross section is much smaller than the parity-conserving one, since the Ar-NO potential is dominated by even $\kappa$-terms, cf. Table~\ref{table:legendre_coefs}. When the field is on, the initial parity is no longer defined. Moreover,  both even and odd Bessel functions $J_{\vert \rho \vert} (k R_0 \vartheta)$ contribute to the cross section. As a result, the differential cross sections corresponding to the final $e$ and $f$ states (dashed curves) become similar to one another.

We see that the field-free differential cross sections, presented in Fig.~\ref{fig:NO-Ar_diff}, are qualitatively similar to the results of close-coupling calculations presented in Fig. 4 of ref.~\cite{Alexander99}, which also show a phase shift between parity-changing and parity conserving cross sections. When the field is turned on, the close coupling calculations also reveal that the cross sections corresponding to final $e$-states exhibit a phase shift and become similar to the cross sections for the final $f$-states, see Fig. 8 of ref.~\cite{Alexander99}.

\section{Conclusions}
\label{sec:conclusions}

We made use of the Fraunhofer model of matter wave scattering to treat rotationally inelastic collisions of polar molecules in electric fields. So far, we have worked out the model for polar molecules in $^1\Sigma$ and $^2\Pi$ states interacting with closed-shell atoms. In accordance with the energy sudden approximation, inherent to the Fraunhofer model, the interaction must be dominated by repulsion. This limits the applicability of the model to the thermal and hyperthermal collision energy range.  
However, the model is also inherently quantum and, therefore, capable of accounting for interference and other non-classical effects. 
The effect of the electric field enters the model via the directional properties of the molecular states and their mixed parity induced by the field. Even a small orientation of the molecule is found to cause a large alteration of the scattering observables, such as differential and integral cross sections. The strength of the analytic model lies in its ability to separate dynamical and geometrical effects and to qualitatively explain the resulting scattering features. These include the angular oscillations in the state-to-state differential cross sections or the rotational-state dependent oscillations in the integral cross sections as a function of the electric field. In the face of the absence of any other analytic model of collisions in fields, the Fraunhofer model is apt at providing a touchstone for understanding such collisions.

\begin{acknowledgements}
We are indebted to Gerard Meijer, Bas van de Meerakker, Steven Hoekstra, Joop Giliamse, and Ludwig Scharfenberg for discussions and encouragement that helped us to get along with the present work. We thank Steven Stolte for his most helpful comments.
\end{acknowledgements}



\newpage

\begin{table}[h]
\centering
\caption{Hard-shell Legendre moments $\Xi_{\kappa 0}$ for the Ne-OCS potential at a collision energy of 500 cm$^{-1}$ and for the Ar-NO potential at a collision energy of 442 cm$^{-1}$.\\}
\label{table:legendre_coefs}
\begin{tabular}{| c | c | c |}
\hline 
\hline
& \multicolumn{2}{| c | }{$\Xi_{\kappa 0}$ (\AA) }   \\[3pt]
\hline
$\kappa$ &  Ne-OCS   & Ar-NO \\[3pt]
\hline
 0 &  14.7043   & 11.0407 \\
 1 & -0.0968     & 0.1744 \\
 2 &  0.9455     & 0.5757 \\
 3 &  0.0540     & 0.0040  \\
 4 & -0.0384     & -0.0713 \\
 5 & -0.0131     & -0.0013 \\
 6 & 0.0012       & 0.0106 \\
 \hline
 \hline
\end{tabular}
\end{table}


\newpage

\end{document}